\documentclass[pre,groupedaddress,reprint,twocolumn,showpacs,notitlepage]{revtex4}

\usepackage{amsfonts}
\usepackage{amsmath}
\usepackage{amssymb}
\usepackage{graphicx}
\usepackage[hidelinks]{hyperref}
\usepackage{color}
\usepackage{mathrsfs}
\usepackage{bbm}
\usepackage{subfigure}
\usepackage{enumerate}
\usepackage{xcolor}
\hypersetup{
    colorlinks,
    linkcolor={blue},
    citecolor={blue},
    urlcolor={blue}
}
\usepackage{times,txfonts}

\usepackage{siunitx}

\newcommand{\ignore}[1]{}
\newcommand{\nobibentry}[1]{{\let\nocite\ignore\bibentry{#1}}}
\newcommand{\bibfnamefont}[1]{#1}
\newcommand{\bibnamefont}[1]{#1}

\newcommand{\ket}[1]{\left\vert#1\right\rangle}
\newcommand{\bra}[1]{\left\langle#1\right\vert}

\newcommand*{\rom}[1]{\expandafter\@slowromancap\romannumeral #1@}
\makeatother

\newcommand{\q}[1]{\dot{\mathcal{Q}}_#1}

\makeatletter
\newcommand\xleftrightarrow[2][]{%
  \ext@arrow 9999{\longleftrightarrowfill@}{#1}{#2}}
\newcommand\longleftrightarrowfill@{%
  \arrowfill@\leftarrow\relbar\rightarrow}
\makeatother

\begin{document}

\title{Internal dissipation and heat leaks in quantum thermodynamic cycles}

\author{Luis A. Correa}
\email{LuisAlberto.Correa@uab.cat}
\affiliation{Departament de F\'{i}sica, Universitat Aut\`{o}noma de Barcelona - E08193 Bellaterra, Spain}

\author{Jos\'{e} P. Palao}
\affiliation{IUdEA Instituto Universitario de Estudios Avanzados, Universidad de La Laguna, 38203 Spain}
\affiliation{Departamento de F\'{i}sica, Universidad de La Laguna, 38204 Spain}

\author{Daniel Alonso}
\affiliation{IUdEA Instituto Universitario de Estudios Avanzados, Universidad de La Laguna, 38203 Spain}
\affiliation{Departamento de F\'{i}sica, Universidad de La Laguna, 38204 Spain}

\date{\today}

\begin{abstract}
The direction of the steady-state heat currents across a generic quantum system connected to multiple baths may be engineered so as to realize virtually any thermodynamic cycle. In spite of their versatility such continuous energy-conversion systems are generally unable to operate at maximum efficiency due to non-negligible sources of irreversible entropy production. In this paper we introduce a minimal model of irreversible absorption chiller. We identify and characterize the different mechanisms responsible for its irreversibility, namely heat leaks and internal dissipation, and gauge their relative impact in the overall cooling performance. We also propose reservoir engineering techniques to minimize these detrimental effects. Finally, by looking into a known three-qubit embodiment of the absorption cooling cycle, we illustrate how our simple model may help to pinpoint the different sources of irreversibility naturally arising in more complex practical heat devices.        
\end{abstract}

\maketitle

\section{Introduction}\label{S_intro}

Almost two centuries ago, Sadi Carnot established the maximum efficiency attainable by heat engines in his groundbreaking work on the motive power of heat \cite{carnot1890fire}. He did so by considering a simple reversible model (the Carnot cycle), which was general enough to account for any specific realization of a steam engine in the limit of \textit{quasistatic} operation. Realistic devices, however, need to be operated in finite time and are, therefore, intrinsically irreversible. Thus, in order to address the practical problem of performance optimization of irreversible engines (or refrigerators), better benchmarks, tighter than the Carnot bound, must be obtained. This is the central question of \textit{finite-time thermodynamics} \cite{chen1999finite}, and addressing it requires to introduce suitable models, simple enough to remain analytically tractable, and still contain all the relevant physics of an irreversible device. 

The first attempts relied on \textit{endoreversible} models, which assume that finite-rate heat transfer effects yield the leading contribution to the total irreversible entropy production, and disregard completely other naturally occurring irreversible processes such as internal dissipation, friction or heat leaks. This approximation, originally intended to account for simple turbine models \cite{chambadal1949thermodynamique,Yvon1955reactor,novikov1957efficiency}, has been generically used in the description of heat engines and refrigerators \cite{hoffmann1997endoreversible,curzon1975efficiency,0022-3727_23_2_002}. However, a careful analysis shows that the operation of commercial devices is indeed dominated instead by internal dissipation and, to a lesser extent, heat leaks, which advises against its use \cite{Gordon2000}. 

Finite-time thermodynamics has also been applied to \textit{quantum} energy conversion cycles, i.e. those embodied in individual quantum systems with discrete energy spectrum \cite{1310.0683v1,e15062100,gelbwaser2015thermodynamics}. These nanoscale heat devices were first studied back in the late 1950s \cite{PhysRevLett.2.262,PhysRev.156.343,5123190}, when it was pointed out that a three-level maser could be thought-of as the smallest realization of a heat engine, or a compression refrigerator. Interestingly, a weakly driven three-level maser is, in good approximation, endoreversible \cite{geva1996quantum}, while a three-level absorption chiller or heat transformer (i.e. driven by heat rather than work) \cite{PhysRevE.64.056130,PhysRevLett.108.070604} is strictly so. The optimization of this model has been thoroughly studied so as to establish universal endoreversible performance bounds from a full microscopic description of the system-baths interactions \cite{esposito2009universality,Correa2013,Correa2014,PhysRevE.89.042128,PhysRevE.90.062124}.

Very recently, several experiments have been proposed to realize quantum energy conversion devices, such as a heat engine made up of a single trapped ion \cite{PhysRevLett.109.203006,PhysRevLett.112.030602}, laser-driven two-level chillers \cite{PhysRevE.87.012140,PhysRevE.87.012120,PhysRevA.91.023431}; or even absorption chillers realized in optomechanical setups \cite{PhysRevLett.108.120602}, arrays of coupled quantum dots \cite{PhysRevLett.110.256801}, or super-conducting qubits \cite{0295-5075_97_4_40003}. Hence, the interest in nanoscale heat devices is starting to transcend the theoretical understanding of the foundations of thermodynamics as their first practical applications start to be envisaged.

Just like in the macroscopic realm, realistic nanometer-sized engines and refrigerators are generally not strictly endoreversible \cite{e15062100}. Therefore, in order to predict characteristic parameters such as the maximum cooling coefficient of performance (COP), or to find generic design prescriptions compatible with operation at maximum power (or cooling rate), one needs to come up with a prototype model, complex enough to encompass all the relevant lossy mechanisms other than the unavoidable finite-rate heat transfer effects. 

For instance, quantum Otto cycles performed \textit{in finite time} on interacting two-level atoms were proposed for the study of `friction' \cite{PhysRevE.65.055102,PhysRevE.68.016101,diosi2006exact,PhysRevE.73.025107}, which was identified with the irreversible degradation of residual quantum coherence. When it comes to continuous models (i.e. those operating under steady-state conditions), it is known that e.g. a weakly-driven three-level chiller does suffer from `heat leaks' once the parasitic coupling between its driving transition and the environment is accounted for \cite{PhysRevE.64.056130}. As shown by Kosloff and Levy in \cite{1310.0683v1}, even perfectly isolated three-level chillers deviate from endoreversibility under strong driving. This is due to an essentially different type of losses, which result from an internal competition between alternative decay paths. This losses are accompanied by heat leaks. Interestingly, both internal losses and heat leaks are also present in absence of external driving or imperfect thermal insulation. Indeed, multi-terminal autonomous devices, i.e. those consisting of mutually interacting contact ports locally addressed by independent heat baths, operate irreversibly as a result of `delocalized dissipation' \cite{Correa2013}. 

In spite of the efforts devoted to the study of irreversibility in nanoscale continuous devices, a complete understanding of its basic underlying mechanisms and their interplay is still missing. In this paper we intend to fill that gap by showing how a generic quantum heat engine or refrigerator can be thought-of as a `multi-effect' device \cite{PhysRevE.89.042128}, made up of \textit{detuned} stages that may work cooperatively and, crucially, also compete with each other \cite{1310.0683v1}. We generically term this competition `internal dissipation'. Heat leaks appear then as a mere by-product, since the combination of such stages facilitates parasitic energy flows between heat baths. Additionally, imperfect insulation at the thermal contacts introduces undesired system-bath couplings that result in a direct bath-to-bath heat transfer or `thermal short-circuit'. To carry out our analysis, we propose a minimal irreversible absorption refrigerator which reduces to a double-stage absorption chiller \cite{PhysRevLett.108.070604}, with detuned three-level stages. We further quantify and compare the individual contributions of internal dissipation and heat leaks to the total irreversible entropy production, and even propose reservoir engineering techniques to partly suppress these detrimental effects and thus, increase the cooling performance. Finally, we show how the basic lossy mechanisms present in our simple model also show up naturally in more complex practical multi-terminal interacting absorption chillers \cite{Correa2013}.

This paper is structured as follows: In Sec.~\ref{S_four_level} we introduce our minimal irreversible model. Its steady-state solution is discussed in detail in Sec.~\ref{S_internal_dissipation_and_heat_leaks}: In particular, we propose a meaningful breakdown of its steady-state heat currents in Secs.~\ref{S_underlying_stochastic_dynamics} and \ref{S_breakdown_of_the_steady_state_heat_currents}. We also discuss the phenomenon of thermal short-circuit in Sec.~\ref{S_thermal_short_circuit}, and propose strategies to partly suppress internal dissipation and heat leaks in Sec.~\ref{S_four_level_supression}. The diagnosis of the more complex three-qubit interacting model is tackled Sec.~\ref{S_three_qubit}. Finally, in Sec.~\ref{S_conclusions} we summarize and draw our conclusions.  

\section{The four-level chiller}\label{S_four_level}

\subsection{Preliminaries: The three-level chiller}\label{S_three_level}

Let us start by looking at the paradigmatic three-level absorption chiller \cite{PhysRevE.64.056130}, with Hamiltonian 
\begin{equation}
H_3=\omega_c\ket{2}\bra{2}+\omega_h\ket{3}\bra{3}.
\label{3_level_hamiltonian}\end{equation}

In all what follows $\hbar=k_B=1$. The transition $\ket{1}\leftrightarrow\ket{2}$, of frequency $\omega_c\,(<\omega_h)$, is weakly coupled to a heat bath at temperature $T_c$ via a dissipative thermal contact term of the form $S_c\otimes B_c\equiv\ket{1}\bra{2}\otimes B_c+\text{h.c.}$. Here $B_c$ is the corresponding bath operator (cf. Sec.~\ref{S_four_level_model}). Likewise, the transitions $\ket{1}\leftrightarrow\ket{3}$ and $\ket{2}\leftrightarrow\ket{3}$, with frequencies $\omega_h$ and $\omega_h-\omega_c$, couple weakly to heat baths at temperatures $T_h$ and $T_w$, respectively. The temperatures are ordered as $T_w>T_h>T_c$.

One may compute the energy currents $\dot{\mathcal{Q}}_{\alpha\,\in\,\{w,h,c\}}$ flowing from each of the heat baths into the system from its (non-equilibrium) steady state $\varrho(\infty)$ \cite{PhysRevE.64.056130}. Provided that 
\begin{equation}
\omega_c < \omega_{c,\text{rev}} = \omega_h\,\frac{T_c (T_w-T_h)}{T_h(T_w-T_c)},
\label{reversiblewc}\end{equation}
one has $\q{w}>0$, $\q{h}<0$ and $\q{c}>0$. That is, net heat is extracted from the cold bath at temperature $T_c$ and dumped into the hot bath at $T_h$, with the assistance of extra energy coming from the work bath at $T_w$. This realizes a \textit{quantum absorption chiller} \cite{PhysRevE.64.056130,PhysRevLett.108.070604}.

Unless $\omega_c=\omega_{c,\text{rev}}$ the operation of the chiller is always irreversible, meaning that the total entropy production $\Delta S\equiv-\q{w}/T_w-\q{h}/T_h-\q{c}/T_c>0$ is strictly positive. This may be understood as a consequence of the `imperfect thermal contact' between each transition and its heat bath. Indeed, let us define the \textit{internal} temperatures $\tau_\alpha$ \cite{PhysRev.156.343,PhysRevE.85.051117} as 
\begin{equation}
\tau_w=-\frac{\omega_h-\omega_c}{\log\left(\frac{p_3}{p_2}\right)}\,,~\tau_h=-\frac{\omega_h}{\log\left(\frac{p_3}{p_1}\right)}\,,~\tau_c=-\frac{\omega_c}{\log\left(\frac{p_2}{p_1}\right)}\,,
\end{equation}
where the $p_i$ stand for the steady-state populations in the energy basis. The equality $\tau_\alpha=T_\alpha$ only holds when $\omega_c=\omega_{c,\text{rev}}$ \cite{PhysRev.156.343}. In that limit $\Delta S=0$ and, as a result, the coefficient of performance $\varepsilon\equiv\q{c}/\q{w}$ of the chiller saturates to the Carnot COP $\varepsilon_C=T_c(T_w-T_h)/[T_w(T_h-T_c)]$ \cite{PhysRevLett.2.262,PhysRev.156.343}. Otherwise, heat transfer is unavoidably irreversible due to the mismatch between $\tau_\alpha$ and $T_\alpha$, and the COP evaluates instead to $\varepsilon=\tau_c(\tau_w-\tau_h)/[\tau_w(\tau_h-\tau_c)]<\varepsilon_C$. 

Unlike the endoreversible three-level maser, the minimal four-level model described below does contain the relevant physical ingredients producing internal dissipation and heat leaks and therefore, it is more akin to a real absorption chiller.      

\subsection{The four-level chiller and its steady-state solution}\label{S_four_level_model}

\begin{figure}
	\includegraphics[width=\columnwidth]{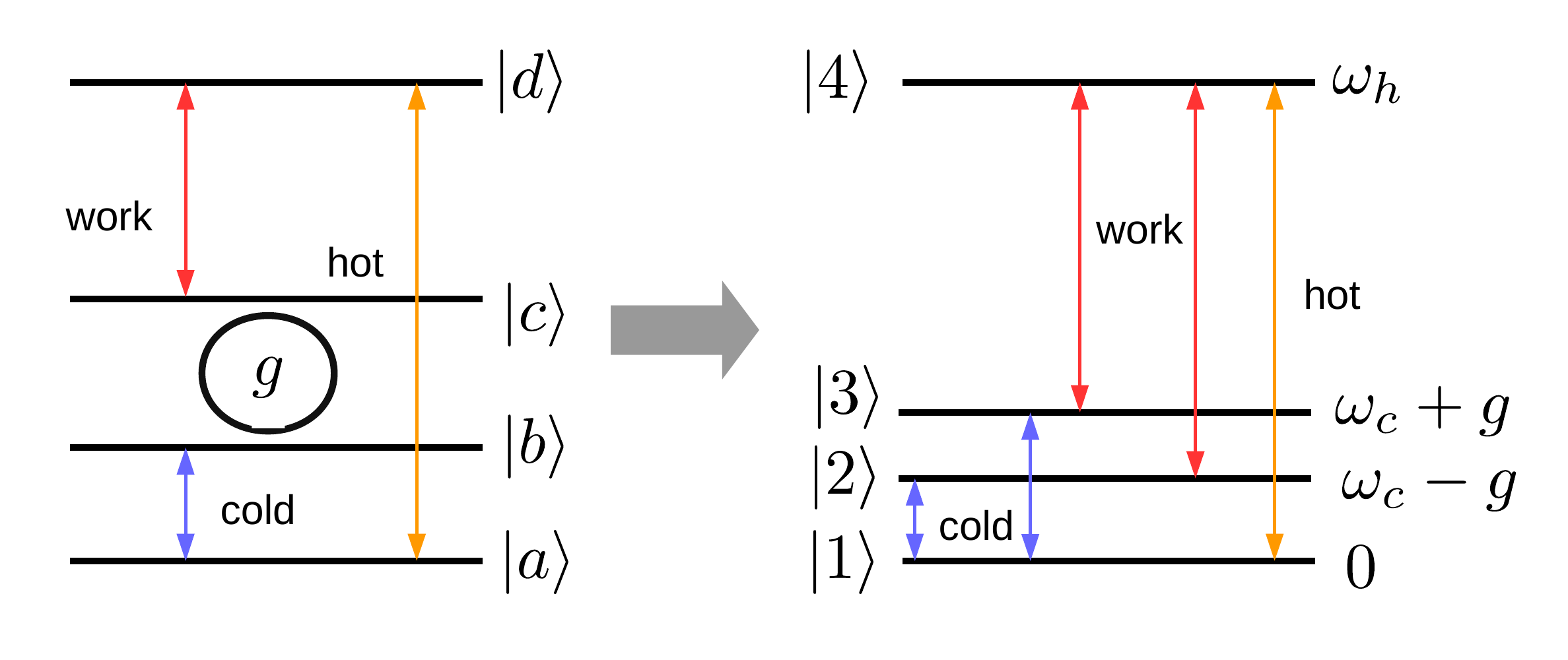}
	\caption{Sketch of the four-level chiller (left). The transitions coupled to each heat bath (work, hot and cold) are indicated by labelled arrows. The $g$-dependent term in $H_4^I$ is depicted by a circle. On the right, the coupling scheme between the energy eigenstates is shown. The corresponding energies are indicated on the right-hand side.}
\label{fig1}
\end{figure}

Let us consider the following Hamiltonian
\begin{multline}
H_4=H_4^{0}+H_4^{I}\equiv\omega_c\left(\ket{b}\bra{b}+\ket{c}\bra{c}\right)+\omega_h\ket{d}\bra{d}\\
+g(\ket{b}\bra{c}+\ket{c}\bra{b}).
\label{4level_hamiltonian}\end{multline}

We couple the transition $\ket{a}\leftrightarrow\ket{b}$ to the cold bath through a thermal contact operator of the form $S_c\otimes B_c\equiv\ket{a}\bra{b}\otimes B_c+\text{h.c}$, the transition $\ket{a}\leftrightarrow\ket{d}$, to the hot bath via a term like $S_h\otimes B_h\equiv\ket{a}\bra{d}\otimes B_h+\text{h.c.}$, and finally $\ket{c}\leftrightarrow\ket{d}$, to the work bath with $S_w\otimes B_w\equiv\ket{c}\bra{d}\otimes B_w+\text{h.c.}$ (see Fig.~\ref{fig1}). In all what follows, we assume $g \ll \omega_h$.

The eigenstates of $H_4$ are $\ket{1}\equiv\ket{a}$, $\ket{2}\equiv 2^{-1/2}(\ket{b}-\ket{c})$, $\ket{3}\equiv 2^{-1/2}(\ket{b}+\ket{c})$, and $\ket{4}\equiv\ket{d}$, and have energies $0$, $\omega_c-g$, $\omega_c+g$, and $\omega_h$, respectively. Henceforth, while the hot bath only acts on the transition $\ket{1}\leftrightarrow\ket{4}$, the work bath is responsible for both the $\ket{2}\leftrightarrow\ket{4}$ and the $\ket{3}\leftrightarrow\ket{4}$ processes. Likewise, the cold bath acts on $\ket{1}\leftrightarrow\ket{2}$ and $\ket{1}\leftrightarrow\ket{3}$. 

We will assume bosonic reservoirs with Hamiltonian $H_{B_\alpha}=\sum_\mu \omega_\mu a_{\alpha\mu}^\dagger a_{\alpha\mu}$ ($\alpha\in\{w,h,c\}$), where $a_{\alpha\mu}$ stands for the annihilation operator of mode $\omega_\mu$ in bath $\alpha$. Hence, we may explicitly write $B_\alpha=\sum_\mu g_\mu (a_{\alpha\mu}+a_{\alpha\mu}^\dagger)$, where the coupling strengths are taken to be $g_\mu=\sqrt{\gamma \omega_\mu}$ so as to yield an Ohmic spectral density \cite{breuer2002theory}. This is the case for e.g. blackbody radiation. The parameter $\gamma$ sets the dissipation time scale $\tau_D\sim\gamma^{-1}$.

We also adopt the usual Born-Markov and rotating-wave approximations, implying that $\tau_D$ is by far the largest time scale in the problem. Under this assumption one may readily derive an equation of motion for the state $\varrho(t)$ in the standard Lindblad-Gorini-Kossakovski-Sudarshan (LGKS) form $\dot\varrho=(\mathcal{L}_w + \mathcal{L}_h + \mathcal{L}_c)\,\varrho$ \cite{lindblad1976generators,gorini1976completely}, where the \textit{dissipators} $\mathcal{L}_\alpha$ are given by
\begin{equation}
\mathcal{L}_\alpha\,\varrho\equiv\sum_{\omega}\Gamma_{\alpha,\,\omega}
\left(A_{\alpha,\,\omega}\,\varrho\,A^\dagger_{\alpha,\,\omega}-\frac{1}{2}\left\lbrace A^\dagger_{\alpha,\,\omega}\,A_{\alpha,\,\omega}\,,\varrho\right\rbrace_+\right),
\label{dissipators}\end{equation}
and $\{\cdot\,,\,\cdot\}_+$ stands for anti commutator. The non-Hermitian jump operators $A_{\alpha,\,\omega}=A^{\dagger}_{\alpha,\,-\omega}$ result from the decomposition of $S_\alpha$ as eigenoperators of $H_4$ \cite{breuer2002theory} (i.e. $S_\alpha\equiv\sum_{\omega}A_{\alpha,\omega}$ such that $[H_4,A_{\alpha,\,\omega}]=-\omega\,A_{\alpha,\,\omega}$). There is only one of them associated with the hot bath ($A_{h,\,\omega_h}=\ket{1}\bra{4}$), which contributes to the corresponding dissipator with two terms, at frequencies $\pm \omega_h$. On the other hand, the work bath adds four terms, at frequencies $\pm(\omega_w\pm g)$, with the associated operators $A_{w,\,\omega_w+g}=\ket{2}\bra{4}$ and $A_{w,\,\omega_w-g}=\ket{3}\bra{4}$. Here, we have defined $\omega_w\equiv\omega_h-\omega_c$. The same happens with the cold bath, that contributes with terms at frequencies $\pm(\omega_c\pm g)$, for which $A_{c,\,\omega_c+g}=\ket{1}\bra{3}$ and $A_{c,\,\omega_c-g}=\ket{1}\bra{2}$.

To ensure thermalization, the relaxation rates $\Gamma_{\alpha,\,\omega}$ must satisfy the detailed balance condition $\Gamma_{\alpha,\,-\omega}/\Gamma_{\alpha,\,\omega}=e^{-\omega/T_\alpha}$. For blackbody radiation these are of the form $\Gamma_{\alpha,\,\omega}=\gamma\omega^3[1+n_\alpha(\omega)]$, where $n_\alpha(\omega)=[\exp{(\omega/T_\alpha)}+1]^{-1}$ is the bosonic occupation number \cite{breuer2002theory}. 

With all the above in mind, one only needs to impose stationarity [i.e. $(\mathcal{L}_w+\mathcal{L}_h+\mathcal{L}_c)\,\varrho(\infty)=0$] to obtain the non-equilibrium steady state $\varrho(\infty)$ of the four-level system. The heat currents can be defined as $\dot{\mathcal{Q}}_\alpha\equiv\text{tr}\{H_4\mathcal{L}_\alpha\varrho(\infty)\}$ \cite{PhysRevE.64.056130}. Obviously, $\sum_\alpha\dot{\mathcal{Q}}_\alpha=0$ as a consequence of the stationarity. Also since the dissipators $\mathcal{L}_\alpha$ are in the LGKS form \cite{lindblad2001non,spohn1978entropy}, one can show that the irreversible entropy production is non-negative ($\Delta S \geq 0$), as should be expected \cite{alicki1979engine}.

We now have at our disposal all the elements to carry out a full thermodynamic analysis of the four-level chiller. We shall use them below for its diagnosis and eventually, for devising a strategy to suppress the internal losses and heat leaks.

\section{Internal dissipation and heat leaks}\label{S_internal_dissipation_and_heat_leaks}

\begin{figure*}
  \includegraphics[width=0.9\columnwidth]{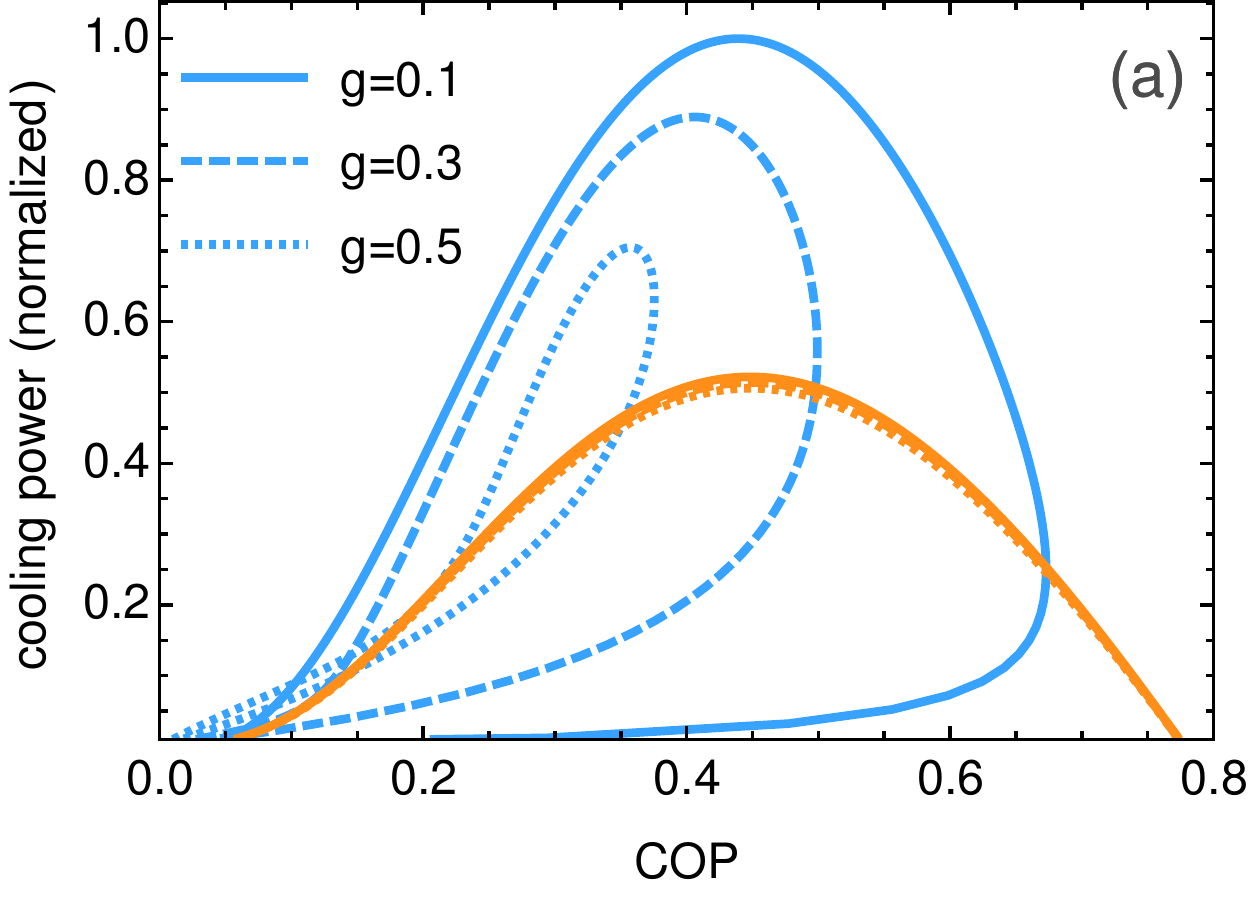}
  \includegraphics[width=0.9\columnwidth]{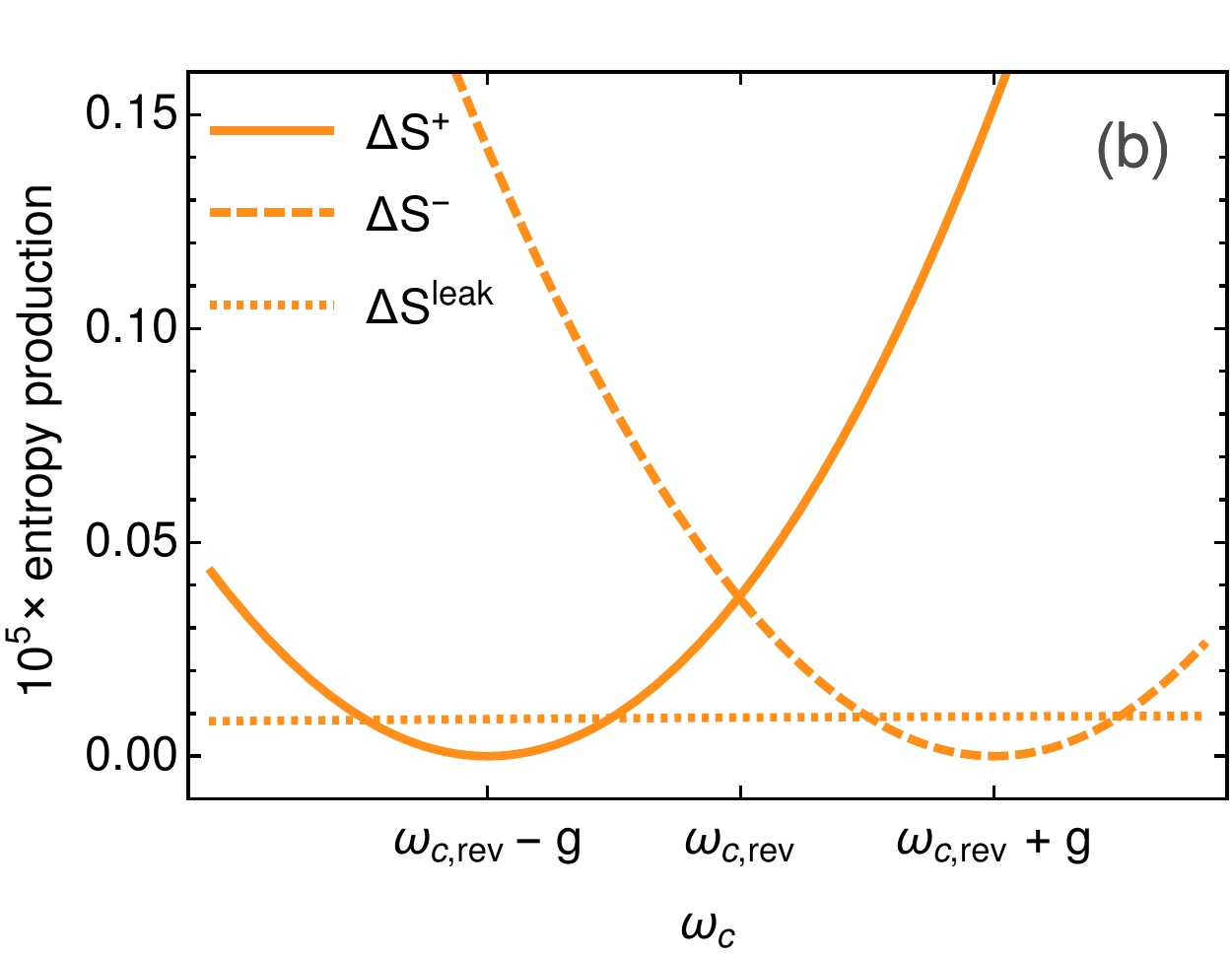}  
  \caption{(a) Characteristic curves of the four-level chiller for $g=\{0.1,0.3,0.5\}$ (\textit{closed} blue curves). The reservoir temperatures are $T_w=9$, $T_h=8$, $T_c=7$, and the hot frequency is set to $\omega_h=6$ (in arbitrary units). To generate these curves, the cold frequency $\omega_c$ was increased until the device stopped cooling. Decoupling the device from all modes of the work bath at frequencies above e.g. $\omega_w=\omega_h-\omega_c$ completely suppresses the losses. The characteristic curves then become \textit{open} like those of endoreversible devices (orange), and the COP saturates to its maximum possible value of $\varepsilon_C=7/9$. (b) Individual contributions $\Delta S^+$ (solid), $\Delta S^-$ (dashed) and $\Delta S^\text{leak}$ (dotted) to the total entropy production $\Delta S=\Delta S^++\Delta S^-+\Delta S^\text{leak}$, in units of $k_B$, around $\omega_{c,\text{rev}}$ for $g=0.1$. The rest of parameters are the same as in panel (a). See discussion in Sec.~\ref{S_breakdown_of_the_steady_state_heat_currents}}
\label{fig2}
\end{figure*}

Heat devices are usually characterized by their power-versus-performance characteristic curves. Endoreversible models feature a distinctive open characteristic curve: When finite-rate heat transfer losses dominate, both the output power and the COP vanish. In the opposite (reversible) limit, the performance saturates to the Carnot COP, albeit at vanishing cooling load [see orange curves in Fig.~\ref{fig2}(a)]. On the contrary, more realistic devices affected as well by internal dissipation and heat leaks exhibit typical closed characteristic curves in which power and efficiency always vanish jointly \cite{gordon1991generalized}. The closed characteristic curves of our four-level absorption chiller, depicted in Fig.~\ref{fig2}(a) in blue, are thus a smoking gun evidence of additional irreversibility. Notice as well that increasing the parameter $g$ seems to accentuate this effect. The aim of this section is to propose a meaningful breakdown of the total heat currents $\q{\alpha}$ so as to identify the sources of the internal dissipation and parasitic heat transfer. 

\subsection{Underlying stochastic dynamics}\label{S_underlying_stochastic_dynamics}

Given a quantum master equation like the one above, one can always tackle the calculation of any physical quantity via the Monte-Carlo wave function (MCWF) method \cite{osa_93}, that is, as an ensemble average over a sufficiently large number of stochastic trajectories. These are generated by a suitable piece-wise deterministic dynamics acting on pure states drawn from a given initial probability distribution. The MCWF method is not only technically convenient for computations involving large dissipative systems \cite{breuer2002theory} but it may also provide physical insights which would be hard to gasp from averaged quantities. 

In particular, the static viewpoint of the system \textit{frozen} at $\varrho(\infty)$ may be replaced by a dynamical one: that of a large number of time-evolving energy eigenstates drawn from the stationary distribution $\{p_1,p_2,p_3,p_4\}\equiv\text{diag}\,\varrho(\infty)$. Note that the state is written here in the energy eigenbasis $\{\ket{1},\ket{2},\ket{3},\ket{4}\}$, rather than $\{\ket{a},\ket{b},\ket{c},\ket{d}\}$. The dynamics along each trajectory would then consist of a sequence of independent random jumps between the different energy eigenstates of the system, at the constant jump rates $\Gamma_{\alpha,\pm\omega_\alpha}$. 

From the specific dissipative couplings of our four-level chiller, as depicted in Fig.~\ref{fig1}(a), it is easy to see that any  \textit{closed} stochastic trajectory (i.e. starting and finishing at the same energy eigenstate), either falls into one of the following six categories, or is a permutation of combinations thereof:
\begin{enumerate}[(C1)]
\item $\ket{1}\xrightarrow[\text{cold}]{\omega_c+g}\ket{3}\xrightarrow[\text{work}]{\omega_w-g}\ket{4}\xrightarrow[\text{hot}]{-\omega_h}\ket{1}$. 
\item $\ket{1}\xrightarrow[\text{hot}]{\omega_h}\ket{4}\xrightarrow[\text{work}]{-(\omega_w-g)}\ket{3}\xrightarrow[\text{cold}]{-(\omega_c+g)}\ket{1}$.
\item $\ket{1}\xrightarrow[\text{cold}]{\omega_c-g}\ket{2}\xrightarrow[\text{work}]{\omega_w+g}\ket{4}\xrightarrow[\text{hot}]{-\omega_h}\ket{1}$.
\item $\ket{1}\xrightarrow[\text{hot}]{\omega_h}\ket{4}\xrightarrow[\text{work}]{-(\omega_w+g)}\ket{2}\xrightarrow[\text{cold}]{-(\omega_c-g)}\ket{1}$.
\item $\ket{1}\xrightarrow[\text{cold}]{\omega_c-g}\ket{2}\xrightarrow[\text{work}]{\omega_w+g}\ket{4}\xrightarrow[\text{work}]{-(\omega_w-g)}\ket{3}\xrightarrow[\text{cold}]{-(\omega_c+g)}\ket{1}$.
\item $\ket{1}\xrightarrow[\text{cold}]{\omega_c+g}\ket{3}\xrightarrow[\text{work}]{\omega_w-g}\ket{4}\xrightarrow[\text{work}]{-(\omega_w+g)}\ket{2}\xrightarrow[\text{cold}]{-(\omega_c-g)}\ket{1}$.
\end{enumerate} 

The energy gained by the four-level system in each jump is indicated above the arrows and the corresponding baths are identified with the labels underneath them. 

Processes C1 and C3 are conventional thermodynamic cooling cycles with a net energy transfer from the cold to the hot bath \cite{PhysRevE.64.056130}, while C2 and C4 are their complementary heat-transforming cycles. On the other hand, the cycles C5 and C6 realize direct energy transfer between the work and the cold baths with no other effect on the hot reservoir, and are thus at the origin of the heat leaks.
When it comes to processes C1/C2 and C3/C4 these may either work cooperatively or compete with each other. For $g<\omega_c<\omega_{c,\text{rev}}-g$ the rates of C1 and C3 are both larger than those of C2 and C4; that is
\begin{subequations}
\begin{align}
\Gamma_{c,\,-(\omega_c+g)}\,\Gamma_{w,\,-(\omega_w-g)}\,\Gamma_{h,\,\omega_h} &> \Gamma_{c,\,\omega_c+g}\,\Gamma_{w,\,\omega_w-g}\,\Gamma_{h,\,-\omega_h} \label{imbalance_plus}\\ 
\Gamma_{c,\,-(\omega_c-g)}\,\Gamma_{w,\,-(\omega_w+g)}\,\Gamma_{h,\,\omega_h} &> \Gamma_{c,\,\omega_c-g}\,\Gamma_{w,\,\omega_w+g}\,\Gamma_{h,\,-\omega_h}. \label{imbalance_minus} 
\end{align}
\end{subequations}

Hence, the two processes cooperate producing net cooling, pretty much like the different stages in a multi-effect quantum absorption chiller \cite{PhysRevE.89.042128}. However, in the range $\omega_{c,\text{rev}}-g < \omega_c < \omega_{c,\text{rev}}+g$, the rate of C2 exceeds that of C1 while C3 is still more likely to occur than C4. This means that the couple C3/C4 cools on average, as opposed to C1/C2, which produces net heating. This competition between cooling and heating mechanisms makes it impossible for the chiller to operate reversibly, and could be tagged `internal dissipation' as it provides the leading contribution to the irreversible entropy production at $\omega_c=\omega_{c,\text{rev}}$ (see Fig.~\ref{fig2}(b) and the discussion in Sec.~\ref{S_breakdown_of_the_steady_state_heat_currents}). For $\omega_c>\omega_{c,\text{rev}}+g$, both C1/C2 and C3/C4 operate as heat transformers on average so that cooling is no longer possible. This is exactly the same mechanism that renders a periodically-driven three-level engine irreversible \cite{1310.0683v1}. 

We shall formalize this intuition below, by suitably decomposing the steady-state heat currents of our four-level chiller into contributions corresponding to two \textit{detuned} elementary endoreversible stages, and an additional purely irreversible heat leak stage operating in parallel.

\subsection{Breakdown of the steady-state heat currents}\label{S_breakdown_of_the_steady_state_heat_currents}

Let us introduce now the steady-state heat currents $\q{w}^{\pm} \equiv (\omega_w\mp g)\,\mathcal{I}_{\pm}$, $\q{h}^{\pm} \equiv -\omega_h\,\mathcal{I}_{\pm}$, $\q{c}^{\pm} \equiv (\omega_c\pm g)\,\mathcal{I}_{\pm}$, and $\q{c}^\text{leak}=-\q{w}^\text{leak}\equiv g\,\mathcal{I}_\text{leak}$, where the rates $\mathcal{I}_{\pm,\text{leak}}$ are given by
\begin{widetext}
\begin{subequations}
\begin{align}
\mathcal{I}_{+} &= \frac{\Gamma_{c,\,\omega_c+g}+\Gamma_{w,\,-(\omega_w-g)}}{D}\left(\Gamma_{c,\,-(\omega_c-g)}\,\Gamma_{w,\,-(\omega_w+g)}\,\Gamma_{h,\,\omega_h}-\Gamma_{c,\,\omega_c-g}\,\Gamma_{w,\,\omega_w+g}\,\Gamma_{h,\,-\omega_h}\right) \label{i_plus} \\
\mathcal{I}_{-} &= \frac{\Gamma_{c,\,\omega_c-g}+\Gamma_{w,\,-(\omega_w+g)}}{D}\left(\Gamma_{c,\,-(\omega_c+g)}\,\Gamma_{w,\,-(\omega_w-g)}\,\Gamma_{h,\,\omega_h}-\Gamma_{c,\,\omega_c+g}\,\Gamma_{w,\,\omega_w-g}\,\Gamma_{h,\,-\omega_h}\right) \label{i_minus} \\
\mathcal{I}_\text{leak} &= \frac{1}{D}\left(\Gamma_{c,\,-(\omega_c+g)}\,\Gamma_{c,\,\omega_c-g}\,\Gamma_{w,\,\omega_w+g}\,\Gamma_{w,\,-(\omega_w-g)}-\Gamma_{c,\,\omega_c+g}\,\Gamma_{c,\,-(\omega_c-g)}\,\Gamma_{w,\,-(\omega_w+g)}\,\Gamma_{w,\,\omega_w-g}\right)\label{i_leak}\\
D &\equiv-\frac{1}{2}\text{det}~
\left(
\begin{array}{cccc}
2 & 2 & 2 & 2 \\
\Gamma_{c,\,-(\omega_c-g)} & -\big(\Gamma_{c,\,\omega_c-g}+\Gamma_{w,\,-(\omega_w+g)}\big) & 0 & \Gamma_{w,\,\omega_w+g} \\
\Gamma_{c,\,-(\omega_c+g)} & 0 & -\big(\Gamma_{c,\,\omega_c+g}+\Gamma_{w,\,-(\omega_w-g)}\big) & \Gamma_{w,\,\omega_w-g} \\
2\Gamma_{h,\,-\omega_h} & \Gamma_{w,\,-(\omega_w+g)} & \Gamma_{w,\,-(\omega_w-g)} & -2\Gamma_{h,\,\omega_h}-\Gamma_{w,\,\omega_w+g}-\Gamma_{w,\,\omega_w-g} 
\end{array}
\right)\,.\label{detD}
\end{align}
\end{subequations}
\end{widetext}

One can easily check that $\q{\alpha}=\q{\alpha}^+ +\q{\alpha}^- + \q{\alpha}^\text{leak}$ for all three baths. The thermodynamic consistency of these newly-introduced \textit{stages} contributing to the total steady-state heat currents, follows from the fact that they individually sum up to zero (i.e. $\sum_\alpha\q{\alpha}^+=\sum_\alpha\q{\alpha}^-=\sum_\alpha\q{\alpha}^\text{leak}=0$), and also satisfy the second-law-like inequalities $\Delta S^+\equiv -\sum_\alpha\q{\alpha}^+/T_\alpha\geq 0$, $\Delta S^-\equiv -\sum_\alpha\q{\alpha}^-/T_\alpha\geq 0$ and $\Delta S^\text{leak}\equiv -\sum_\alpha\q{\alpha}^\text{leak}/T_\alpha\geq 0$.

We may associate the stage $\{\,\q{\alpha}^+\,\}$ to the combined processes C1/C2, as the sign of $\mathcal{I}_+$ is determined by the imbalance between the stationary rates of processes C1 and C2 [cf. Eqs.~\eqref{i_plus} and \eqref{imbalance_plus}]. Similarly, the sets of steady-state heat currents $\{\,\q{\alpha}^-\,\}$ and $\{\,\q{\alpha}^\text{leak}\,\}$ may be linked to processes C3/C4 and C5/C6, respectively. 

The `plus stage' behaves as an absorption chiller (i.e. $\q{w}^+>0$, $\q{h}^+<0$ and $\q{c}^+>0$) within the \textit{cooling window} $0 < \omega_c < \omega_{c,\text{rev}}-g$, while the `minus stage' does, provided that $g < \omega_c <\omega_{c,\text{rev}}+g$ \cite{Correa2014}. On the other hand, the heat leak rate $\mathcal{I}_\text{leak}$ is always negative since $T_w>T_c$. This entails a parasitic heat current flowing directly from the work bath into the cold bath. Since we set $g\ll\omega_h$, the heat leaks will be typically negligible in comparison with the other contributions. Also note that as $g$ increases, the cooling window for the plus stage shrinks while, for the minus stage, it is instead shifted as a whole towards larger frequencies. Thus, if we were to allow $g$ to become larger than $\omega_{c,\text{rev}}$, the plus stage would not be capable of producing net cooling at \textit{any} $\omega_c$, and the four-level chiller could ultimately lose its ability to cool.

Just like in the case of the three-level chiller, the internal temperatures $\tau_w^+$, $\tau_h^+$ and $\tau_c^+$, defined from the steady-state populations $\{p_1,p_3,p_4\}$ in the energy basis as
\begin{equation}
\tau_w^+=-\frac{\omega_w-g}{\log\left(\frac{p_4}{p_3}\right)}\,,~\tau_h^+=-\frac{\omega_h}{\log\left(\frac{p_4}{p_1}\right)}\,,~\tau_c^+=-\frac{\omega_c+g}{\log\left(\frac{p_3}{p_1}\right)}\,,
\end{equation} 
coincide with the equilibrium temperatures of the reservoirs only at $\omega_c=\omega_{c,\text{rev}}-g\equiv\omega_{c,\text{rev}}^+$. At that point, the plus stage operates reversibly ($\Delta S^+=0$), i.e. at the Carnot COP. Similarly, defining the internal temperatures $\tau_\alpha^-$ from $\{p_1,p_2,p_4\}$ one can see that the equalities $\tau_\alpha^-=T_\alpha$ and $\Delta S^-=0$ hold at $\omega_c=\omega_{c,\text{rev}}+g\equiv\omega_{c,\text{rev}}^-$. However, the irreversible entropy production associated to the heat leak stage $\Delta S^\text{leak}$ never vanishes. That would require both $\tau_c^\pm=T_c$ \textit{and} $\tau_w^\pm=T_w$ to hold simultaneously, which is not possible for any $g\neq 0$.  

Fig.~\ref{fig2}(b) shows the behaviour of the different shares of the total irreversible entropy production ($\Delta S^+$, $\Delta S^-$ and $\Delta S^\text{leak}$) in the neighbourhood of $\omega_c=\omega_{c,\text{rev}}$. We can see that even if a non-vanishing $\Delta S^\text{leak}$ alone is enough to render the device overall irreversible, the leading detrimental effect comes from the mismatch between $\omega_{c,\text{rev}}^+$ and $\omega_{c,\text{rev}}^-$. Also note that the larger the $g$, the larger the mismatch and, consequently, the bigger the resulting irreversibility at $\omega_{c,\text{rev}}$. This explains why increasing $g$ lowers the maximum attainable COP in Fig.~\ref{fig2}(a).    

\begin{figure}
	\includegraphics[width=\columnwidth]{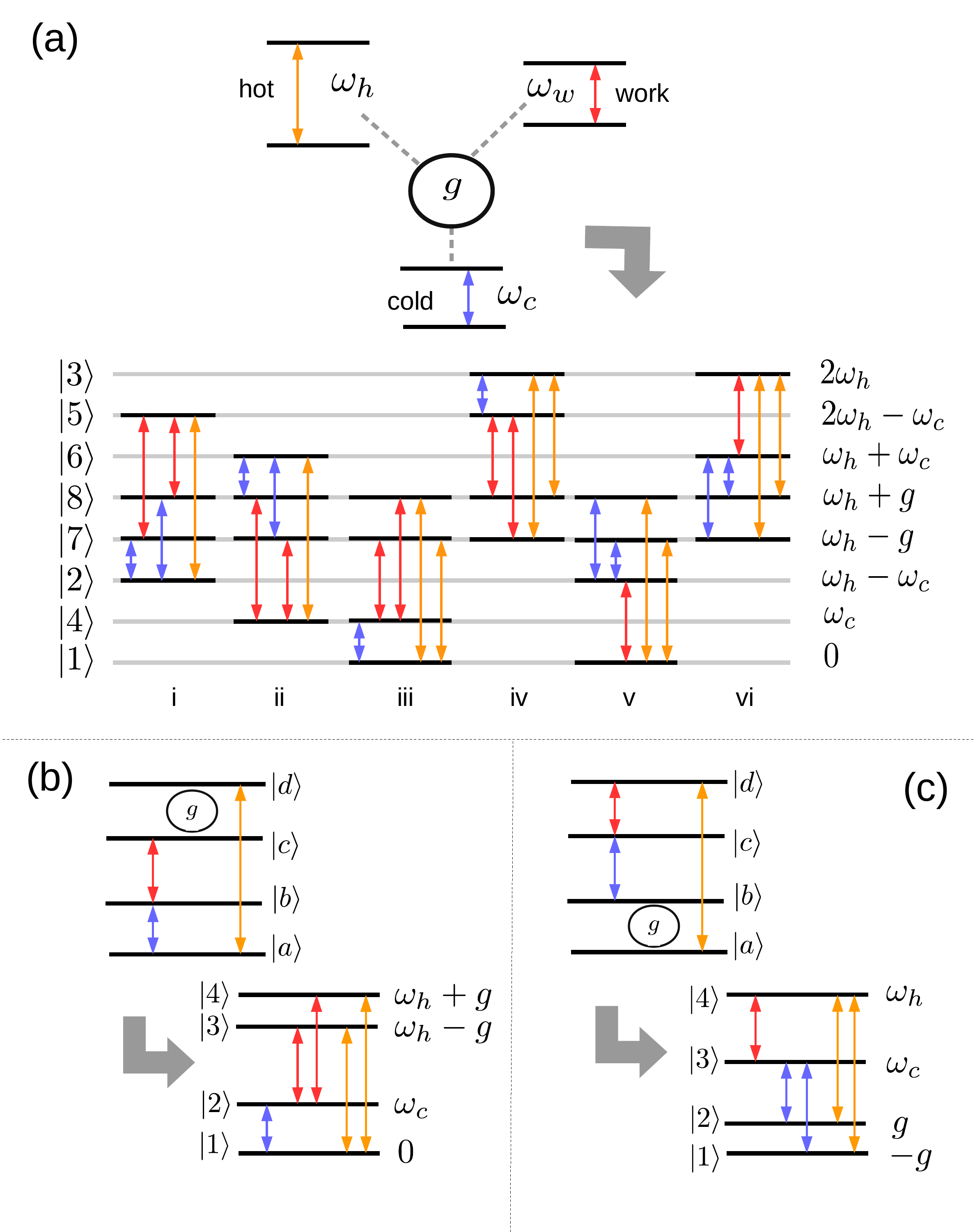}
	\caption{(a) Sketch of the three-qubit absorption chiller and its transition diagram. The \textit{eigen-energies} are shown on the right, while the kets on the left denote the energy eigenstates following the notation of Ref.~\cite{Correa2014}. The transitions driven by the work, hot and cold baths are indicated with red, orange and blue arrows, respectively (see text for details). (b and c) Sketch of the two alternative configurations of the four-level chiller with their corresponding transition diagrams (see text for details).}
\label{fig3}
\end{figure}

\subsection{Thermal short circuit: A different type of heat leaks}\label{S_thermal_short_circuit}

Generalizing the above, we can say that a thermodynamic cooling cycle will feature irreversibility provided that two (or more) energy-basis transitions at different frequencies are allowed by the coupling to two (or more) heat baths. The dissipative transitions of such system may always be grouped to form coupled detuned three-level stages, which gives rise to \textit{both} internal dissipation and heat leaks as illustrated above.

An essentially different scenario is discussed in Ref.~\cite{PhysRevE.64.056130}, where irreversibility was modeled as the parasitic coupling of e.g. the cold bath to the work transition in a three-level maser. This would provide the chiller with an alternative dissipative path besides the conventional cooling and heating cycles $\ket{1}\xleftrightarrow{\text{cold}}\ket{2}\xleftrightarrow{\text{work}}\ket{3}\xleftrightarrow{\text{hot}}\ket{1}$; that is, the \textit{shortcut} $\ket{2}\xleftrightarrow{\text{work}}\ket{3}\xleftrightarrow{\text{cold}}\ket{2}$. As a result, heat could leak between the work and cold bath without producing any cooling. While the former effect is mere a by-product of internal dissipation, rooted in the interactions facilitating energy transfer within the working substance, the origin of this latter type of heat leaks, which we may tag `thermal short circuit', lies simply in the lack of proper insulation at the thermal contacts of the device. Realizable nanoscale absorption chillers will typically be multi-terminal interacting devices, thus suffering from internal dissipation and heat leaks. However, thermal short circuit effects may also be unavoidable in practice, and could play a major role in the buildup of irreversible entropy production. 
  
\subsection{Suppression of internal dissipation and heat leaks}\label{S_four_level_supression}

From the above analysis it follows that the overall operation of the four-level chiller can be made endoreversible by inhibiting either the plus or the minus stage. Let us imagine for instance that the spectrum of the work bath could be tailored into a Debye shape, with a high frequency cutoff $\omega_{w,\text{max}}$ anywhere between $\omega_w-g$ and $\omega_w+g$. The corresponding excitation-relaxation rates would then read $\tilde{\Gamma}_{w,\,\omega}=\gamma\omega^3\,[1+n_w(\omega)]\,\Theta(\omega_{w,\text{max}}-\vert\omega\vert)$, where $\Theta(\cdots)$ stands for the Heaviside step function. As a result, the stationary rates associated with the transition $\ket{2}\leftrightarrow\ket{4}$ would vanish, thus precluding processes C3/C4 and C5/C6, and cancelling their contributions to the total irreversible entropy production. Such engineered four-level chiller can operate reversibly at $\omega_c=\omega_{c,\text{rev}}^+$, as showcased by its open characteristic curve, plotted in orange in Fig.~\ref{fig2}(a).

It is important to stress that the Markov approximation, which is central to our analysis, assumes that the decay of the two-time correlations of the baths is sufficiently fast compared with the dissipation time-scale. This assumption is justified so long as the spectral density remains approximately flat in the neighbourhood of the relevant Bohr frequencies of the system. Thus, for small $g$, the cutoff would be typically too close to $\omega_w-g$ for the Markov approximation to be valid. As a result, both the actual time-evolution and the subsequent stationary state might differ significantly from the ones predicted by the LGKS quantum master equation. A rigorous treatment of such problem requires more sophisticated tools, like higher order perturbative quantum master equations \cite{breuer2001time,breuer2006nonmarkov}, non-Markovian stochastic Schr\"{o}dinger equations \cite{diosi1997non,diosi1998non,Vega2005}, hierarchical Heisenberg equations \cite{PhysRevLett.94.200403,PhysRevA.75.052108}, or other non-perturbative methods \cite{Imamoglu1994,PhysRevA.55.2290,gelbwaser2015strongly}. 

While the quantitative assessment of the performance of irreversible devices operating between engineered reservoirs is definitely worthy of investigation, it lies beyond the scope of this paper. Nevertheless, from the simple physical picture sketched above one may still propose practical means to minimize the detrimental effects of heat leaks and internal dissipation. For instance, rather than connecting the transition $\ket{c}\leftrightarrow\ket{d}$ directly to the work bath, one could interpose a harmonic oscillator of frequency $\omega_f$ between the two, so as to realize a frequency filter: The spectral density would then effectively look like a \textit{skewed} Lorentzian around $\omega_f$ \cite{kofman1994spontaneous}. This reservoir-engineering technique has been recently considered in the context of nanoscale heat devices to produce spectrally-separated heat baths \cite{PhysRevE.87.012140,gelbwaser2014heat}. Adjusting the detuning $\omega_w-\omega_f$ and the transition-oscillator coupling, one can easily find instances in which $\tilde{\Gamma}_{w,\,\omega_w-g}/\tilde{\Gamma}_{w,\,\omega_w+g}\lll 1$, which strongly reduces the rate of the minus stage in favour of the plus stage, and may thus allow for significantly larger cooling COPs.

\section{The three-qubit chiller}\label{S_three_qubit}

We now turn our attention to a more complex three-terminal absorption chiller \cite{PhysRevLett.105.130401,PhysRevLett.108.070604}. These models are known to operate irreversibly as they feature closed characteristic curves \cite{Correa2013,1310.0683v1}. In particular, we consider three interacting two-level atoms, each one coupled to a heat bath, again at temperatures $T_w>T_h>T_c$. As already mentioned, there exist experimental proposals for the implementation of such systems on superconducting qubits \cite{0295-5075_97_4_40003} or quantum dots \cite{PhysRevLett.110.256801}. 

More specifically, the model Hamiltonian reads
\begin{multline}
H_8=\omega_w\ket{1_w}\bra{1_w}+\omega_h\ket{1_h}\bra{1_h}+\omega_c\ket{1_c}\bra{1_c}\\
+g\left(\ket{1_w 0_h 1_c}\bra{0_w 1_h 0_c}+\text{h.c.}\right),
\end{multline} 
in the eight-dimensional three-qubit Hilbert state space $\mathcal{H}_w\otimes\mathcal{H}_h\otimes\mathcal{H}_c$, where $\ket{0_\alpha}$ stands for the ground state and $\ket{1_\alpha}$, for the excited state \cite{PhysRevLett.105.130401} [see Fig.~\ref{fig3}(a)]. The work frequency is set by the resonance condition $\omega_w=\omega_h-\omega_c$. The system-bath couplings may be written as $\sum_\alpha\sigma_x^{(\alpha)}\otimes B_\alpha$, where $\sigma_x^{(\alpha)}$ is the Pauli $x$-matrix in $\mathcal{H}_\alpha$. Details on the derivation of the LGKS quantum master equation for this model can be found in Ref.~\cite{Correa2013}. 

It is crucial to remember that even if the terminals are \textit{locally} addressed by independent baths, the resulting dissipators are \textit{global} due to the three-body interaction term. It is precisely the ensuing `delocalized dissipation' that ultimately makes the device irreversible \cite{Correa2013}. This hints that heat leaks could be at least partly responsible for the irreversibility. As pointed out in Ref.~\cite{levy2014local}, an incorrect local modelling of the dissipation would not only fail to capture relevant physics of the problem, but might even predict physically unacceptable solutions which violate the second law of thermodynamics. 

\subsection{Breakdown of the three-qubit chiller}\label{S_three_qubit_suppression}

In Fig.~\ref{fig3}(a) all the dissipative transitions between the energy eigenstates of the three-qubit chiller are depicted with coloured arrows. Red stands for transitions coupled to the work bath, orange for those coupled to the hot bath, and blue for those of the cold bath. This device features nine open decay channels allowing for a total of eighteen dissipative transitions between the eight energy eigenstates. Even if the dissipative dynamics of such model may seem quite involved at first glance, we will illustrate in what follows how our qualitative understanding of the four-level chiller can greatly help in the system diagnosis and the subsequent performance optimization. 

\subsubsection{Variations of the four-level chiller}

To proceed further, we must discuss first the two possible variations of the irreversible four-level chiller. The first one is illustrated in Fig.~\ref{fig3}(b). Its Hamiltonian reads
\begin{equation}
H'_4=\omega_c\ket{b}\bra{b}+\omega_h(\ket{c}\bra{c}+\ket{d}\bra{d})+g(\ket{c}\bra{d}+\ket{d}\bra{c}),
\end{equation}
while the system-bath coupling operators are $S_c=\ket{a}\bra{b}+\ket{b}\bra{a}$, $S_h=\ket{a}\bra{d}+\ket{d}\bra{a}$, and $S_w=\ket{b}\bra{c}+\ket{c}\bra{b}$. Repeating the analysis from Sec.~\ref{S_internal_dissipation_and_heat_leaks} for this alternative model, one finds that the competition between two detuned endoreversible stages is still responsible for most of the irreversible entropy production, the main difference being that the parasitic heat current leaks from the work bath into the hot bath, rather than into the cold bath. On the other hand, the model depicted in Fig.~\ref{fig3}(c) corresponds to the Hamiltonian
\begin{equation}
H''_4=\omega_c\ket{c}\bra{c}+\omega_h\ket{d}\bra{d}+g(\ket{a}\bra{b}+\ket{b}\bra{a}),
\end{equation}
and system-bath coupling operators $S_c=\ket{b}\bra{c}+\ket{c}\bra{b}$, $S_h=\ket{a}\bra{d}+\ket{d}\bra{a}$, and $S_w=\ket{c}\bra{d}+\ket{d}\bra{c}$. In this configuration the chiller features heat leaks from the hot bath into the cold bath instead.

\subsubsection{Diagnosis of the three-qubit chiller}

Generically, the dissipative transitions of any absorption chiller may be grouped into (possibly detuned) coupled three-level stages. We know from Sec.~\ref{S_breakdown_of_the_steady_state_heat_currents} that the mismatch in the cooling windows of these stages gives rise to internal dissipation, and that the frequency difference between pairs of transitions coupled to different baths facilitates heat leaks. One may easily identify the direction of the leaks by just \textit{pairing off} these elementary three-level stages to yield either of the three configuration of the irreversible four-level chiller. In particular, notice that in Fig.~\ref{fig3}(a) we have split the dissipative transitions of the three-qubit chiller into six coupled irreversible cycles, labelled with lowercase Roman numerals. One recognizes the transition diagram of Fig.~\ref{fig1} in stages (i) and (ii). In turn, stages (iii) and (iv) correspond to the configuration of Fig.~\ref{fig3}(b), and stages (v) and (vi), to that of Fig.~\ref{fig3}(c). Hence, we can conclude that this device suffers from all three types of pairwise heat leaks. However, for small $g$, it is again the mismatch in the cooling windows of the detuned three-level stages which contributes with the largest share to the total irreversible entropy production.

Performance optimization would demand, in this case, to engineer the coupling to both the work and the hot baths. For instance, setting high-frequency cutoffs anywhere between $\omega_h-g$ and $\omega_h$ in the spectrum of the hot bath, and between $\omega_w-g$ and $\omega_w$ in the spectrum of the work bath, suppresses all three-level stages except for the two that connect states $\{\ket{3}$, $\ket{5}, \ket{8}\}$, and $\{\ket{1}, \ket{4}, \ket{7}\}$. These are resonant and uncoupled, and the resulting device is therefore strictly endoreversible.

We have thus showcased how one may diagnose a complex autonomous heat device, and even propose generic strategies to optimize its performance, without ever having to solve for its steady-state heat currents.  

\subsection{COP at maximum power of irreversible devices}\label{S_performance_optimization}

The ultimate efficiency at maximum power that can be achieved by nanoscale heat devices has been object of intense study \cite{geva1991spin,esposito2009universality,schmiedl2008efficiency,PhysRevLett.111.050601,PhysRevE.81.051129}. In particular, for endoreversible single and multi-stage devices it has been established that the COP at maximum cooling load ($\varepsilon_*$) is \textit{tightly} bounded from above as a result of the constraints imposed by the specific system-bath interaction mechanism \cite{PhysRevE.90.062124,PhysRevE.89.042128}. In particular, for unstructured bosonic baths, this bound is only set by the dimensionality of the heat reservoirs (e.g. $\varepsilon_* \leq \frac{3}{4}\varepsilon_C$ for 3D baths) \cite{Correa2014}. Whether the performance bound also applies to irreversible (`non-ideal') chiller models such as the three-qubit chiller was investigated via global numerical optimization in Ref.~\cite{Correa2013}. We are now in the position to give simple arguments to justify why the bound overestimates the maximum performance for any finite internal interaction strength.

For simplicity, we turn our attention to the four-level chiller, though similar arguments apply as well to the three-qubit model. As pointed out in Sec.~\ref{S_breakdown_of_the_steady_state_heat_currents}, the contribution of the heat leak component to the total steady-state heat currents and irreversible entropy production is rather small, so that one may approximate $\q{\alpha}\simeq\q{\alpha}^++\q{\alpha}^{-}$. In particular, the power of each stage $\q{c}^\pm$ will peak at a different frequency $\omega_{c,*}^\pm$, and feature a COP of $\varepsilon_*^\pm=\q{c}^\pm(\omega_{c,*}^\pm)/\q{w}^\pm(\omega_{c,*}^\pm)\leq\frac{3}{4}\varepsilon_C$. However, the total cooling load $\q{c}$ will peak at some $\omega_{c,*}\neq\omega_{c,*}^\pm$, so that one has the strict inequalities $\varepsilon^\pm(\omega_{c,*})<\frac{3}{4}\varepsilon_C$. The overall COP at maximum cooling power may be thus approximated as
\begin{equation}
\varepsilon_*\simeq\frac{\q{w}^+(\omega_{c,*})}{\q{w}(\omega_{c,*})}\varepsilon_+(\omega_{c,*}) + \frac{\q{w}^-(\omega_{c,*})}{\q{w}(\omega_{c,*})}\varepsilon_-(\omega_{c,*}) < \frac{3}{4}\varepsilon_C,
\label{COP_max_pow}\end{equation}
which is \textit{strictly below} the endoreversible bound for unstructured baths. In fact, is it easy to check that increasing the interaction strength $g$ gradually lowers the actual bound on $\varepsilon_*$. Of course, if one is allowed to optimize $\varepsilon_*$ globally (i.e. over all parameters), the endoreversible bound will tighten for vanishingly small interactions $g$. 

This is a neat example of why one should be careful not to extend results obtained within the endoreversible approximation to generic irreversible devices. Especial care is in order when dealing with performance optimization, as endoreversibility typically leads to gross overestimations of the COP.

\section{Conclusions}\label{S_conclusions}

In this paper we have identified and characterized the mechanisms producing internal dissipation and heat leaks in nanoscale autonomous energy-conversion devices. These effects are severely detrimental to their energy-efficient operation and relate closely to each other. They arise whenever it becomes possible for one or more heat baths to excite two or more transitions of the working substance, at \textit{different} frequencies. To that end, we have introduced and solved a minimal model of four-level irreversible absorption chiller. In particular, we have split its steady-state heat currents into three contributions: two of them which correspond to coupled endoreversible three-level stages operating in parallel, and another one which facilitates direct heat leaks from one heat bath into another. We have shown that, since the two endoreversible stages are detuned, there is a range of parameters for which they compete with each other instead of working cooperatively. It is precisely the mismatch in the cooling windows of these two stages which ultimately produces the largest share to the total irreversible entropy production as the coefficient of performance grows. We have generically termed this effect `internal dissipation' to differentiate it from the mere bath-to-bath parasitic heat leaks. We have also proposed the suppression of either of the two constituent three-level cycles of this irreversible model, via reservoir-engineering techniques, as a means to increase its maximum achievable coefficient of performance. Finally, we have shown how, aided by this qualitative understanding about the origin of irreversibility in autonomous heat devices, one may be able to classify the sources of irreversible entropy production of more complex models of absorption chiller, e.g. determine the direction of the heat leaks; and even to develop an intuition about potentially successful reservoir engineering techniques, to be implemented so as to allow for larger performance.

A quantitative measure of the actual improvement in chiller performance achieved, for instance, by introducing fine-tuned frequency filters at the thermal contacts, calls for a more careful modelling of the dissipative interactions beyond the common Markovian approximation. It is certainly in order to pursue this research direction further, as it could ultimately pave the way towards a theory of thermal engineering at the nanoscale, with foreseeable widespread applications to quantum technologies.

\acknowledgments

The authors would like to thank D. Gelbwaser-Klimovsky for useful advice. 
Financial support from EU Collaborative Project TherMiQ (Grant Agreement 618074), Spanish MINECO (through projects FIS2013-40627-P and FIS2013-41352-P), and COST Action MP1209 is gratefully acknowledged.

\bibliographystyle{apsrev}

\end{document}